\documentclass[11pt,a4paper]{article}
\pdfoutput=1
\usepackage{jcappub}
\usepackage{graphicx}

\usepackage{graphicx}
\usepackage{subcaption}
\usepackage{amsmath,amssymb}
\usepackage[font=small,labelfont=bf]{caption}
\usepackage{verbatim}
 \usepackage{url}

\def\araa{\ref@jnl{ARA\&A}}

\title{Observing patchy reionization with future CMB polarization experiments}
\author[a]{A. Roy,}
\author[a,b,c]{A. Lapi,}
\author[d,e]{D. Spergel,}
\author[a,b,c]{C. Baccigalupi}
\affiliation[a]{SISSA, Via Bonomea 265, 34136 Trieste, Italy}
\affiliation[b]{INFN-Sezione di Trieste, via Valerio 2, 34127 Trieste, Italy}
\affiliation[c]{INAF-Osservatorio Astronomico di Trieste, via Tiepolo 11, 34131 Trieste, Italy}
\affiliation[d]{Department of Astrophysical Sciences, Princeton University, Princeton, NJ 08544, USA}
\affiliation[e]{Center for Computational Astrophysics, Flatiron Institute, 162 5th Ave, New York, NY 10003, USA}
\emailAdd{aroy@sissa.it, lapi@sissa.it, dns@astro.princeton.edu, bacci@sissa.it}

\abstract{We study the signal from patchy reionization in view of the future high accuracy polarization measurements of the Cosmic Microwave Background (CMB). We implement an extraction procedure of the patchy reionization signal analogous to CMB lensing. We evaluate the signal to noise ratio (SNR) for the future Stage IV (S4) CMB experiment. The signal has a broad peak centered on the degree angular scales, with a long tail at higher multipoles. The CMB S4 experiment can effectively constrain the properties of reionization by measuring the signal on degree scales. The signal amplitude depends on the properties of the structure determining the reionization morphology. We describe bubbles having radii distributed log-normally. The expected S/N is sensitive to the mean bubble radius: $\bar{R}=5$ Mpc implies $S/N \approx 4$, $\bar{R}=10$ Mpc implies $S/N \approx 20$. The spread of the radii distribution strongly affects the integrated SNR, that changes by a factor of $10^2$ when $\sigma_{lnr}$ goes from $\ln 2$ to $\ln3$. Future CMB experiments will thus place important constraints on the physics of reionization. }


\begin{document}
\maketitle
\flushbottom
\section{Introduction}\label{sec|intro}

Ultraviolet radiation from first sources ionizes the Inter-Galactic Medium (IGM) and alters the thermal, ionization and chemical properties of the gas. The reionization optical
depth $\tau$, weighting Thomson cross section along the line of sight, represents a most important characterization
of the effect. Through the measurements of large-scale $E$ mode polarization in the Cosmic Microwave Background
(CMB) anisotropies ("reionization bump"), the Wilkinson Microwave Anisotropy Probe (WMAP) has constrained the
sky average optical depth to be $0.089\pm 0.014$ \citep{2013ApJS..208...19H}; more recently, the Planck
satellite, by combining data ranging from 30 to 353 GHz in polarization, reported a lower value,
$\tau= 0.058\pm 0.012$ \citep{2016A&A...596A.108P}.

Accurate $\tau$ measurements are crucial for breaking degeneracies with the amplitude of the primordial scalar perturbations $A_s$ in order to infer the dependence of the primordial power with physical scales. However, a precise astrophysical knowledge of the reionization process is necessary for interpreting and exploiting the $\tau$ measurements expected from future CMB polarization experiments. Most important targets of these efforts are represented by the $B$-modes of CMB polarization from cosmological Gravitational Waves (GWs) and Gravitational Lensing (GL). Currently the POLARBEAR/Simons Array \citep{Polarbear2014, Simonsarray2016}, BICEP\footnote{Background Imaging of Cosmic Extragalactic Polarization, \tt{www.cfa.harvard.edu/CMB/bicep3}}3 \citep{BICEP32014}, ACTpol\footnote{Atacama Cosmology Telescope (polarization sensitive), \tt{https://act.princeton.edu/}}\citep{ACTpol2017}, SPT 3G\footnote{South Pole Telescope (third generation), \tt{https://pole.uchicago.edu/}}\citep{SPT3G2014}, SPIDER\footnote{\tt{https://spider.princeton.edu}} \citep{SPIDER2010}, EBEX\footnote{The E and B Experiment, \tt{http://groups.physics.umn.edu/cosmology/ebex/}} \citep{EBEX2010} and others\footnote{See {\tt lambda.gsfc.nasa.gov} for a complete list of operating and planned CMB experiments.} are searching these signals. In the near future the Simons Observatory \footnote{\tt https://simonsobservatory.org} will be observing from the Atacama desert, paving the way to a network of ground based systems equipped with $10^4$ detectors, which will represent the Stage-IV (S4) phase of ground based CMB experiment.

The detection of almost complete Ly$\alpha$ absorption in the spectra of distant quasars may suggest that the end of reionization occurred around $z\approx 6$ \citep{Fan2006}. Recent data on the evolution of Ly$\alpha$ optical depth show a sharp increase for redshift $z$ larger than $6-7$. A similar conclusion is supported by the study of line of sight variations in the IGM Ly$\alpha$ optical depth \citep{Becker2015}. Moreover, the Ly$\alpha$ transmission profile of the highest-redshift quasar (Quasi-Stellar Objects, QSO) ULAS $J112001.48 +064124.3$ at $z\sim 7.1$ is strikingly different from that of two lower redshift $z\sim 6.3-6.4$  counterparts detected in the Sloan Digital Sky Survey (SDSS\footnote{{\tt www.sdss.org}}). It features a measured near-zone radius of $\sim1.9$ Mpc, a factor of $\sim 3$ smaller than it is typical for QSOs at $z \sim 6-6.5$ \citep{Mortlock2011}, suggesting higher$-z$ QSOs live in an IGM whose HI fraction is much higher. High redshift GRBs with their bright afterglow can be also exploited as probes of cosmic reionization, dispensing with some of the complications inherent to QSO observations. Extreme drop off in transmission profile of GRB140515A at z $\approx 6.3$  places an upper limit on HI fraction at these redshift  \citep{Chornock2013, Totani2014}. Several studies support the picture that first galaxies in the Universe are the main source of reionization \citep{Barkana2001, Robertson2010, Lapi2017} but the morphology of the reionization process is still poorly understood. Numerical simulations as well as analytical studies suggest that ionization fraction is spatially inhomogeneous \citep{Furlanetto2005bubble, Barkana2001, Lidz2006}. Reionization sources basically first ionize the surrounding IGM by producing ionized bubbles around them and later they grow and merge with each other \citep{Furlanetto2005, Paranjape2014}. Patchy reionization produces different scattering histories along different line of sights so that the value of $\tau$ varies with the direction. Patchy reionization also generate kinetic Sunyaev Zeldovich (kSZ) signal due to the peculiar motion of ionized bubbles along the line of sight \citep{Natarajan2013, Iliev2006}. In the near future it will be possible to detect patchy kSZ signal and separating that from other secondary anisotropies in the CMB, by exploiting an accurate astrophysical modelling \citep{Smith2017}. \citet{Namikawa2018} constrained the optical depth fluctuations by estimating the trispectrum from Planck 2015 CMB temeperature anistropies data.

In this paper we study the capability of future CMB experiments concerning the reconstruction and understanding
of the reionization process, by probing not only the sky averaged reionization hystory, but also the accessible information regarding its morphology through the dependence on the line of sight. That would provide invaluable insight into the reionization process which would open unprecedented windows on the astrophysical processes responsible for the EoR, but also allowing to reconstruct the overall behavior of cosmology at those epochs.

The paper is organized as follows: in Section \ref{sec|reion} we discuss a realistic model of reionization based on recent determination of the star formation rate functions. In Section \ref{reio_cmb} we study the effects of reionization on the CMB. In Section \ref{reconstruction} we implement the reconstruction method first introduced by \citet{Dvorkin2009} and forecast the capabilities of future CMB experiments with specifications corresponding to S4 to detect patchy reionization. In Section \ref{forecast} we summarize our findings.  Throughout this work we assume flat $\Lambda CDM$ cosmology with parameters $h=0.677$, $\Omega_bh^2=0.02230$, $\Omega_ch^2=0.1188$, $\Omega_\Lambda=0.6911$, $n_s=0.9667$, $A_s=2.142\times 10^{-9}$, derived from the combinations of Planck TT,TE,EE,LowP+Lensing+Ext \citep{Planck2015}. Stellar masses and luminosities of galaxies are based on a Chabrier initial mass function \citep{2003PASP..115..763C}.

\section{Reionization history}\label{sec|reion}

In this section we describe our model for the reionization history. We base on the assumption that high-redshift star-forming galaxies are the primary source of ionizing photons. The cosmic ionization history is basically determined from the cosmic star formation history, in the form
\begin{equation}
\dot N_{\rm ion}\approx f_{\rm esc}\, k_{\rm ion}\, \rho_{\rm SFR}~;
\end{equation}
here $k_{\rm ion}\approx 4\times 10^{53}$ is the number of ionizing photons s$^{-1}$ $(M_\odot/\rm yr)^{-1}$ per unit time and SFR, with the quoted value appropriate for a Chabrier IMF; $f_{\rm esc}\approx 10\%$ is the (poorly constrained) average escape fraction for ionizing photons from the interstellar medium of high-redshift galaxies \citep[see][]{Mao2007, Dunlop2013, Robertson2015, Lapi2017}; and $\rho_{\rm SFR}$ is the cosmic star formation density.

For $\rho_{\rm SFR}$ we assume the determination by \citet{Lapi2017} (see also \citet{Madau2014}), to which we defer the reader for details. This is based on the joint analyses of recent dust-corrected UV \citep[e.g., ][]{Bouwens2015, Bouwens2016}, far-IR \citep[e.g., ][]{Lapi2011, Gruppioni2013, Gruppioni2015, Rowan2016} and radio \citep[e.g., ][]{Novak2017} luminosity functions down to UV magnitudes $M_{\rm UV}\lesssim -17$ out to $z\lesssim 10$ via blank field surveys, and pushed down to $M_{\rm UV} \approx -13$ at $z\lesssim 6$ via gravitational lensing by foreground galaxy clusters \citep[see][]{Alavi2014, Alavi2016, Livermore2017, Bouwens2016}. The resulting $\rho_{\rm SFR}$ significantly depends on the fainter UV magnitude $M_{\rm UV}^{\rm lim}$ (or smaller SFR after $M_{\rm UV}\approx -18.5-2.5\, \log SFR [M_\odot~{\rm yr}^{-1}]$) considered to contribute to the ionizing background.

The competition between ionization and recombination determines the evolution of the ionization state of the universe \citep[see][]{Madau1999, Ferrara2014}:
\begin{equation}
\dot Q_{\rm HII} = {\dot N_{\rm ion}\over \bar n_{\rm H}}-{Q_{\rm HII}\over
t_{\rm rec}}
\label{ionfrac}
\end{equation}
Here $\bar n_{\rm H}\approx 2\times 10^{-7}\, (\Omega_b h^2/0.022)$ cm$^{-3}$ is the mean comoving hydrogen number density. In addition, the recombination timescale reads $t_{\rm rec}\approx 3.2$ Gyr $[(1+z)/7]^{-3}\, C_{\rm HII}^{-1}$, where the case B coefficient for an IGM temperature of $2\times 10^4$ K has been used; this timescale crucially depends on the clumping factor of the ionized hydrogen, for which a fiducial value $C_{\rm HII}\approx 3$ is usually adopted \citep[see][]{Pawlik2013}.

The electron scattering optical depth is proportional to the integrated electron density along the line-of-sight:
\begin{equation}\label{eq_tau_eerm}
\tau_{\rm es}(z) = c\, \sigma_{\rm T}\,\bar n_{\rm H}\int_t^{t_0}{\rm d}t'\,f_e\,Q_{\rm HII}(z')(1+z')^2 ;
\end{equation}
here $dt=\frac{dz}{H(z)}$ and $H(z)=H_0\,[\Omega_M\,(1+z)^3+1-\Omega_M]^{1/2}$ is the Hubble
parameter, $c$ is the speed of light, $\sigma_{\rm T}$ the Thomson cross
section and $f_e$ the number of free-electrons per ionized hydrogen atom (assuming double Helium ionization at $z\lesssim 4$.).

Figure~\ref{fig|reion} shows the reionization history computed from our cosmic SFR density integrated down to different UV magnitude limits $M_{\rm UV}^{\rm lim}$, on assuming a standard value $f_{\rm esc}\approx 0.1$ for the escape fraction of ionizing photons. When adopting $M_{\rm UV}^{\rm lim}\approx -13$, the outcome (black dot-dashed line) agrees with the value of the optical depth for electron scattering $\tau_{\rm }\approx 0.058$ recently measured by the \textsl{Planck} mission. For reference, the dotted line represents the optical depth expected in a fully ionized Universe up to redshift $z$; this is to show that the bulk of the reionization process occurred at $z \sim 8-9$ and was almost completed at $z \sim 6$ \citep[see][]{Schultz2014}. {The thin lines in the same figure show the outcome when the SFR density is truncated for $z>8$. Note that from these perspectives, the detailed behavior of the cosmic SFR density at $z\lesssim 6$ and its extrapolation beyond at $z\gtrsim 10$ are only marginally relevant.

When adopting $M_{\rm UV}^{\rm lim}\approx -17$, that corresponds to the observational limits of current blank-field UV surveys at $z\gtrsim 6$, the outcome on the optical depth (black solid line) touches the lower boundary of the 1$\sigma$ region allowed by \textsl{Planck} data. At the other end, going much beyond $M_{\rm UV}^{\rm lim}\approx -13$ is not allowed, since already for $M_{\rm UV}^{\rm lim}\approx -12$ the resulting optical depth (black dashed line) touches the upper boundary of the 1$\sigma$ region from \textsl{Planck} data.

\begin{figure}[t]
\begin{center}
\includegraphics[width=1\textwidth]{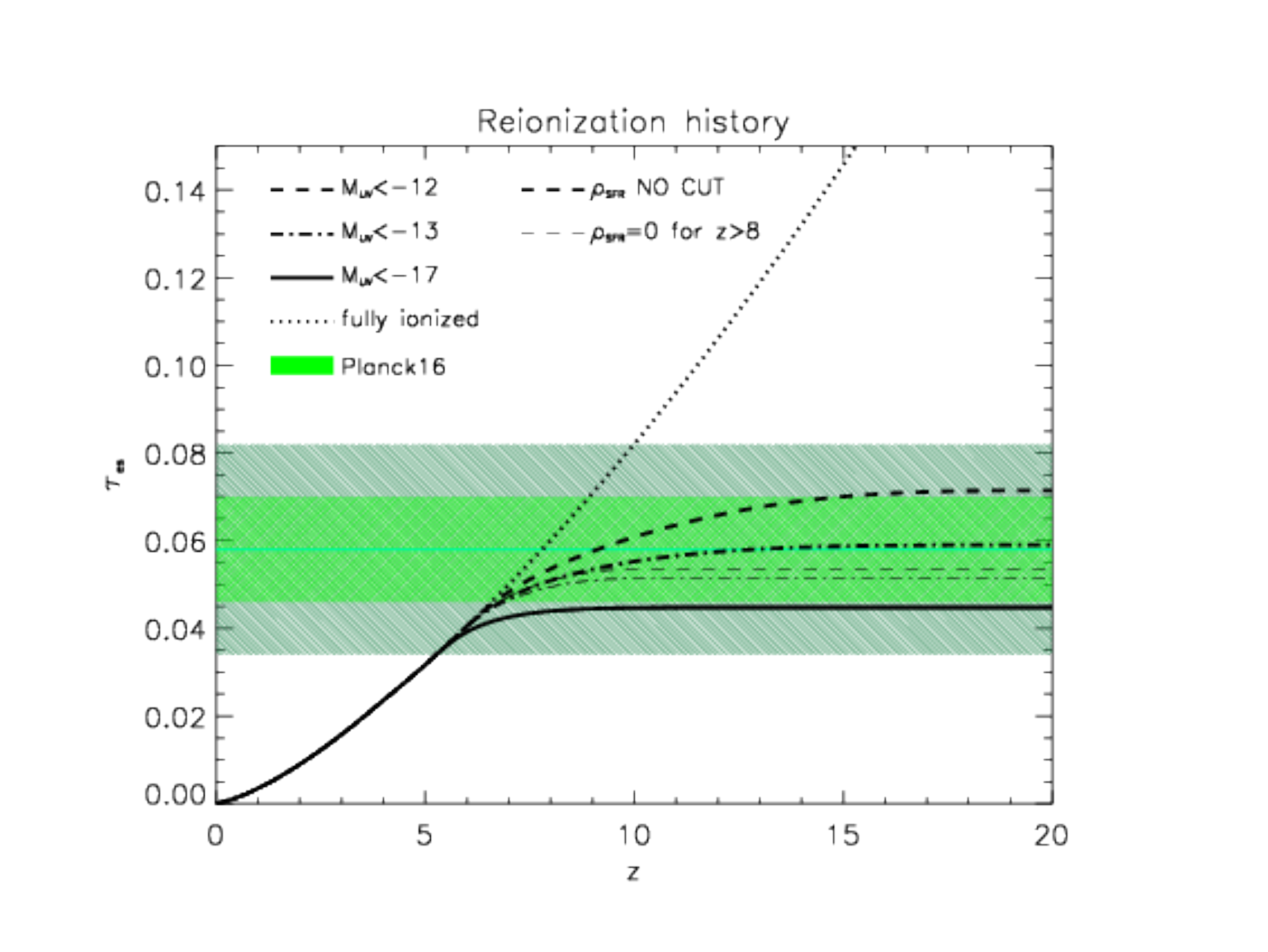}
\caption{Reionization history of the universe, in terms of the redshift
evolution of the optical depth $\tau_{\rm es}$ for electron scattering. Thick solid, dot-dashed and dashed lines illustrate the outcomes of the EERM for the cosmic SFR density integrated down to UV-magnitude limits $M_{\rm UV}\lesssim -17$, $-13$, and $-12$, respectively; thin lines refer to SFR density truncated at $z>8$. For reference, the dotted line refers to a fully ionized universe up to redshift $z$. The green line shows the measurement (with the $1\sigma$ and $2\sigma$ uncertainty regions highlighted by the dark and light green areas respectively) from \citet{Planckxlvii}.}\label{fig|reion}
\end{center}
\end{figure}

\begin{table}[t]
\begin{center}
    \begin{tabular}{| c | c | c | c | c |}
    \hline
    Experiment & Sensitivity $\Delta_T$ & $\theta_f$ \\
    name  & [$\mu$K arcminute]& [arcminute] \\ \hline
    S4 (a) &1 & 1 \\ \hline
    S4 (b)  & 0.5 & 1 \\ \hline
    Litebird & 1.8 & 16 \\
    \hline
    \end{tabular}
    \caption{$\theta_f$ is the full width half maxima of the incident beam. S4 (a) case is the configuration taken from the CMB S4 first science book \citep{CMBS42016} but as it is not still built up, just to be optimistic we are considering higher sensitive configuration in S4 (b) case. }
    \label{tt}
\end{center}
\end{table}

\section{CMB effects from Reionization}\label{reio_cmb}
The cosmological reionization process makes its own signatures in the observed CMB sky. If reionization process did not take place, CMB photons could free-stream preserving their spectral shape originated at the time of recombination. The CMB spectral shape changes due to the y-type distortion by Comptonization process during reionization epoch \citep{sunyaevrishi2013}. The Compton y parameter is of the order of $1.93\times 10^{-7}$ for $\tau=0.058$ and electron temperature $T_e=2\times 10^4$ K \citep{dezotti2015}. Rich groups at $z\sim 1$ will produce a larger y distortion signal, $\sim 1-2\times 10^{-6}$ \citep{Hill2015}. Due to the generation of free electrons along the line of sight, CMB photons scattered off by those electrons during the reionization epoch, and Thomson scattering generates new polarization at the large angular scales which adds up with the polarization induced at the last scattering surface.  Primary CMB anisotropies are damped in all scales by a factor $e^{-\bar{\tau}}$, where $\bar{\tau}$ is the mean reionization optical depth across the sky. Similar to CMB lensing, reionization also creates non-Gaussianity on CMB by correlating different Fourier modes.  In this section we will discuss reionization morphology effects on the CMB by taking into account the reionization model discussed in the previous section and hereafter labelled as Empirical Extended Reionization Model (EERM). Throughout this paper we adopt two foreseen experiments representing the constraining capability of the CMB in the next decade. The first is represented by the network of ground based observatories mounting $\sim 10^4$ detectors, known as stage-IV experiment (S4) and the second one is the LiteBIRD satellite from the Japanese space agency, optimized for large scale CMB polarization observations \citep{Matsumura2014}. The main features of the probes in terms of angular resolution and sensitivity are summarized in Table~\ref{tt}.

\subsection{Sky-averaged effect}

The sky-averaged reionization optical depth has been defined in Eq. (\ref{eq_tau_eerm}) and can be also written as
\begin{equation}\label{tauz_eq}
\tau(z)=c\sigma_T\int_0^z \bar{n}_e(z^\prime) d z^\prime \frac{dt}{dz^\prime},
\end{equation}
Where $c$ is the velocity of light in free space, $\sigma_T$ is the Thomson scattering cross section and $\bar{n}_e(z)$ is the mean free electron number density. As in Eq. (\ref{eq_tau_eerm}) we express $\bar{n}_e(z)$ in terms of present number density of protons $n_{p0}$ and ionization fraction $x_e(z)$ so that $\bar{n}_e(z)= n_{p0}(1+z)^3x_e(z)$ and $dt/dz=[H(z)(1+z)]^{-1}$, where $H(z)$ is the Hubble parameter.

The tanh reionization model \citep{Lewis2008} is most commonly used to parametrize the reionization history of the universe and it is implemented inside Boltzmann equation solver codes like CAMB , CLASS \citep{Lewis2008,classcode}:
\begin{equation}\label{xe_eq}
x_e(z)=\frac{1}{2}\left[1+\tanh\left(\frac{y_{re}-y(z)}{\Delta_y}\right)\right],
\end{equation}
where $y(z)=(1+z)^{\frac{3}{2}}$ ,  $y_{re}=y(z_{re})$ and $\Delta_y=1.5\sqrt{(1+z)}\Delta_z$. Two free parameters of the model are the redshift $z_{re}$ at which ionization becomes at the 50\% level and $\Delta_z$, the width of the transition from neutral to fully ionizing state in redshift units.

\begin{figure}[t]
\begin{center}
\includegraphics[width=1\textwidth]{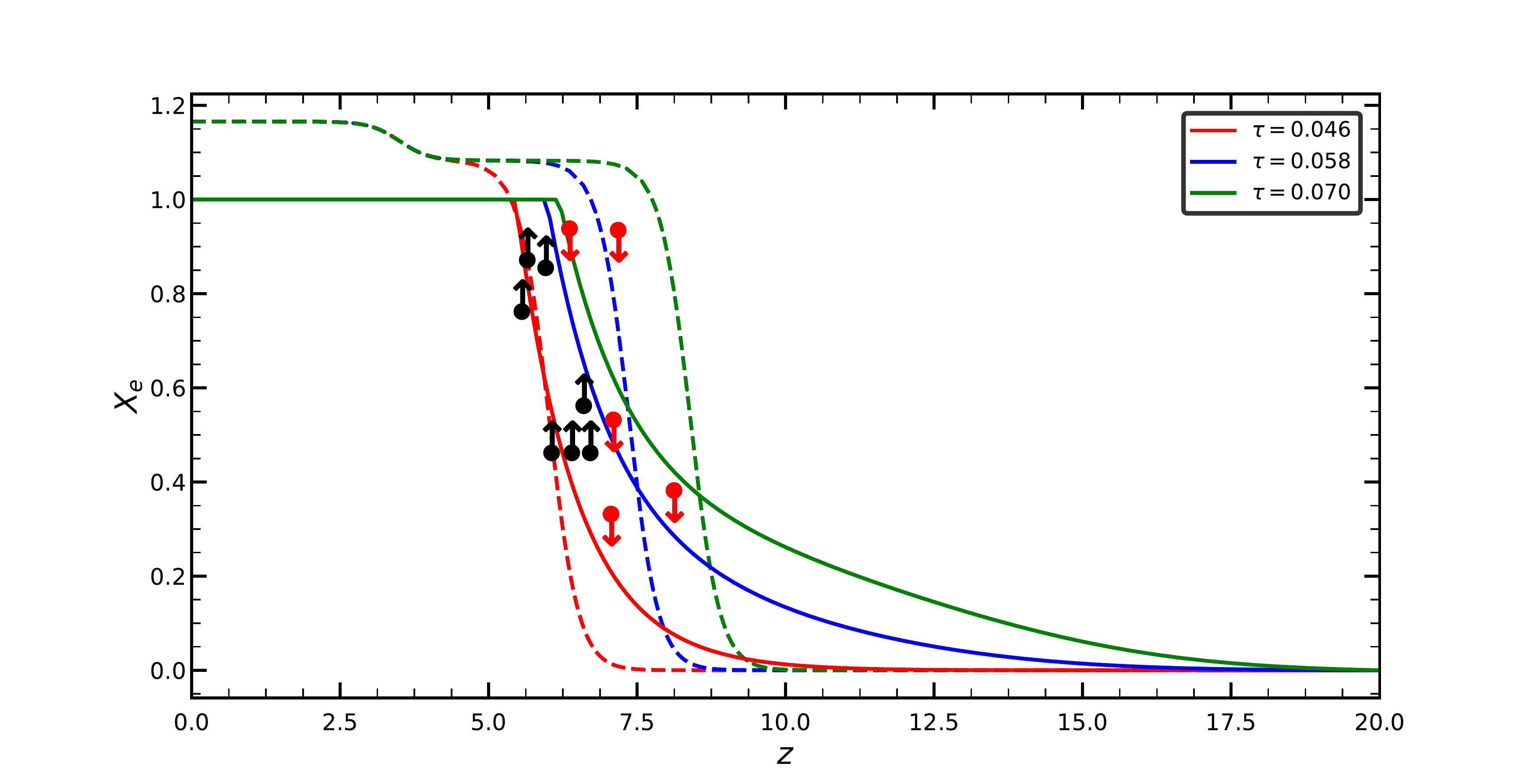}
\caption{Solid lines represent the ionization history of the universe for three different optical depths considering EERM; the dashed lines are reionization histories calculated by the CLASS code using the tanh model for the same optical depths. Upper (red) and lower (black) observational limits from various data collected by \cite{Robertson2015}.}
\label{fig:plot_xe}
\end{center}
\end{figure}

We implement our reionization model, EERM by modifying CLASS code to check to which extent the EE and BB power spectrum sensitive to different ionization histories. In Figure~\ref{fig:plot_xe} we compare the ionization history for EERM and the tanh reionization model corresponding to $\tau=0.070$ (green), $0.058$ (blue) and $0.044$ (red),  together with upper and lower limits from various observations collected by \citet[][empty circles]{Robertson2015}. EERM predicts more extended reionization process roughly in between the redshift range 10 to 6 corresponding to $\tau=0.058$ (solid blue line) whereas for the tanh case $x_e$ changes more sharply at around $z\approx 7-8$  (dashed blue line). In Figure~\ref{fig:clee} EE spectra show the effect of the optical depth $\tau$ as the amplitude of the bump increases with higher $\tau$ value. Size of the horizon at the time of reionization is much larger than the horizon size at the time of last scattering, hence the $\ell$ range at which the EE spectra become maximum is sensitive to mean redshift of reionization. The height of the reionization bump in the EE spectra is maximum around $\ell \approx 4$ for EERM whereas for tanh reionization model it is around $\ell \approx 3$. The shift is due to a slight anticipation of the reionization process in the tanh model. However, the difference between EERM ad the tanh model is difficult to probe at the power spectrum level, as shown by the errorbars on the EE spectra for
the LiteBIRD sensitivity.

\begin{figure}[h!]
\begin{center}
\includegraphics[width=1\textwidth]{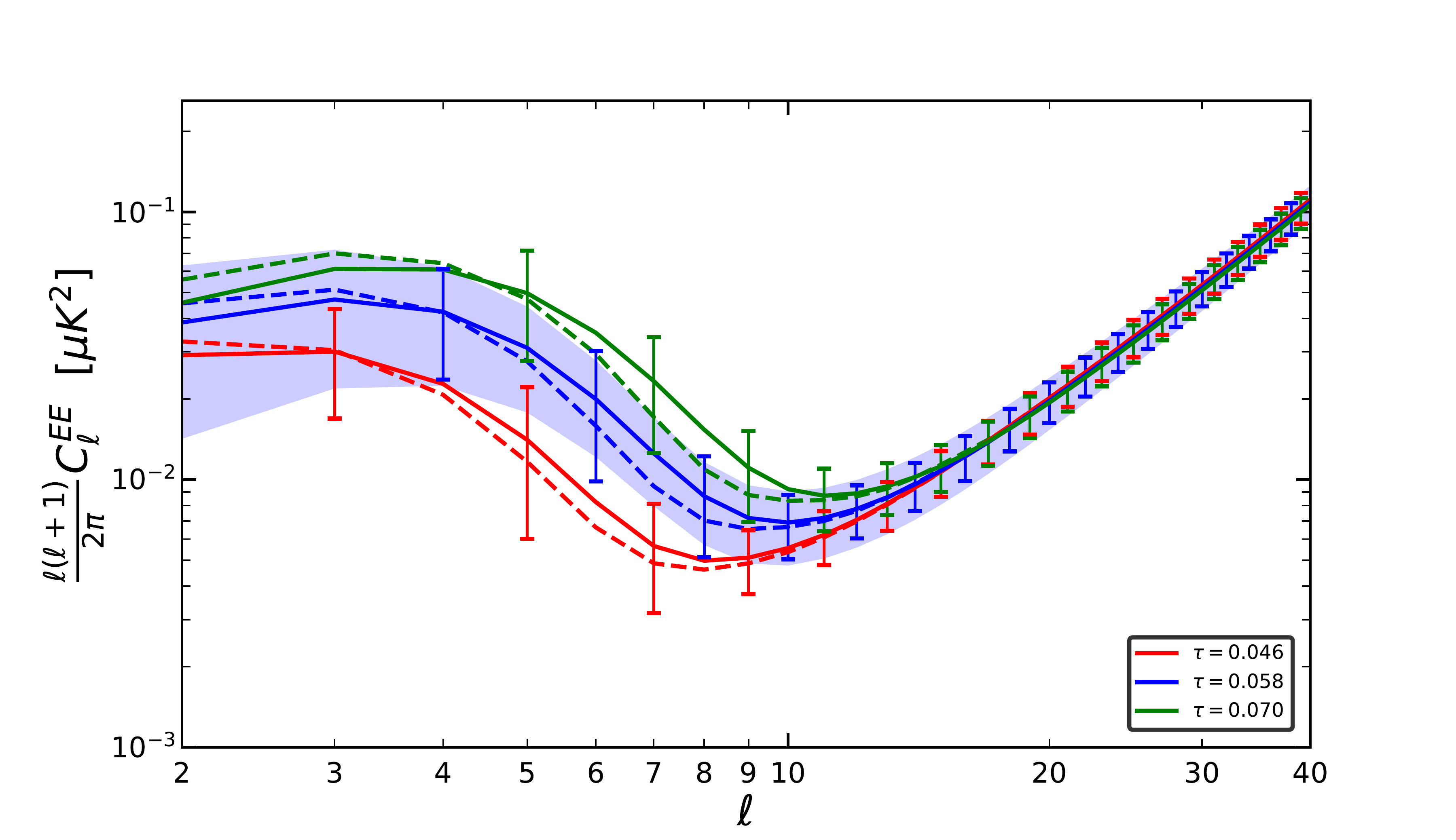}
\includegraphics[width=1\textwidth]{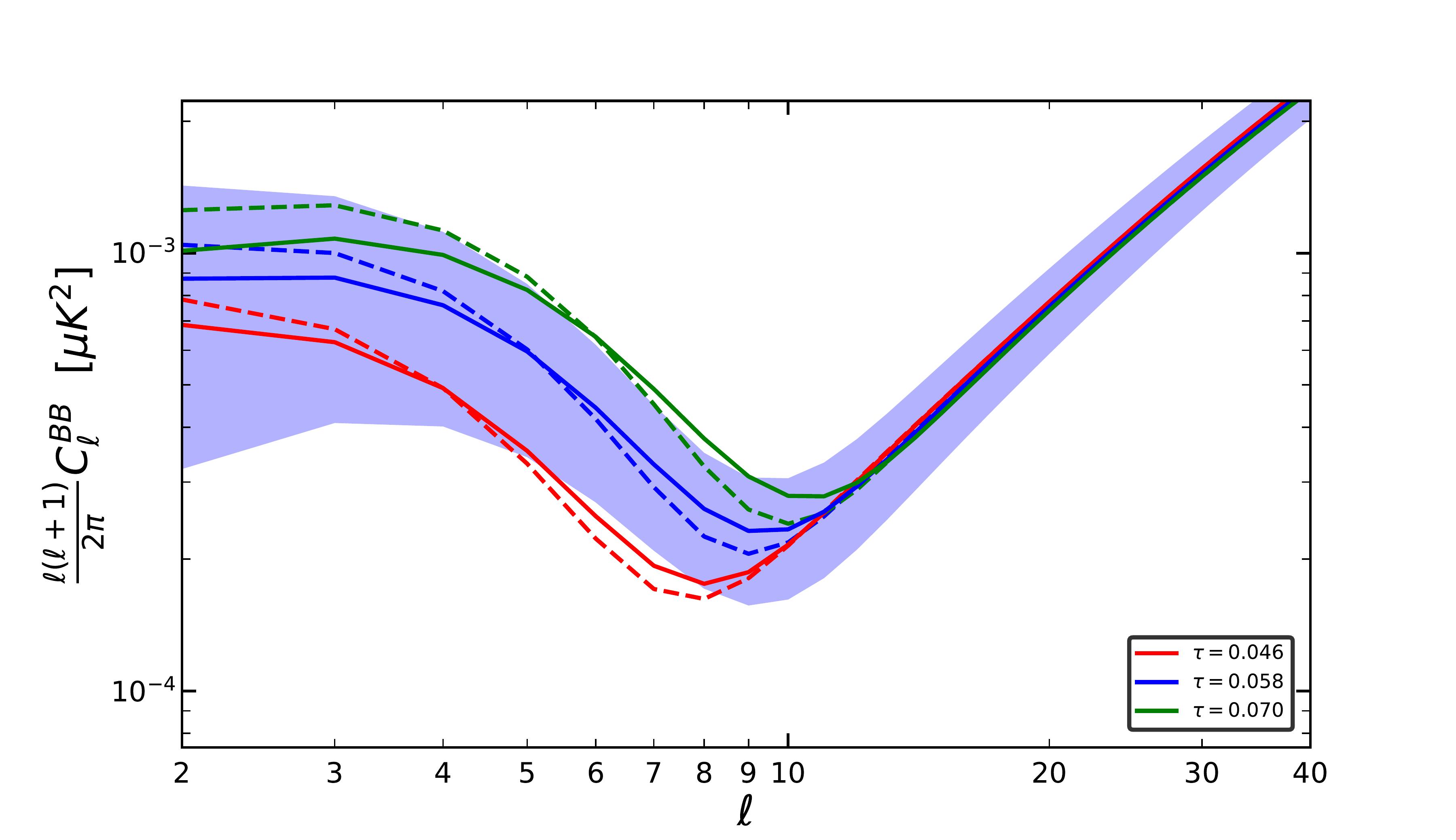}
		
\caption{Upper panel: solid lines describe the EE power spectra for EERM and dashed lines for the tanh model, on adopting  $\tau= 0.046$ (red), $0.058$ (blue) and $0.070$ (green). Shaded area shows the 1$\sigma$ cosmic variance limit corresponding to $\tau=0.058$. Errorbars refer to the LiteBIRD sensitivity.  In the bottom panel we plot the corresponding BB spectra. }
\label{fig:clee}
\end{center}
\end{figure}

\begin{figure}[h!]
\begin{center}
\includegraphics[width=1\textwidth]{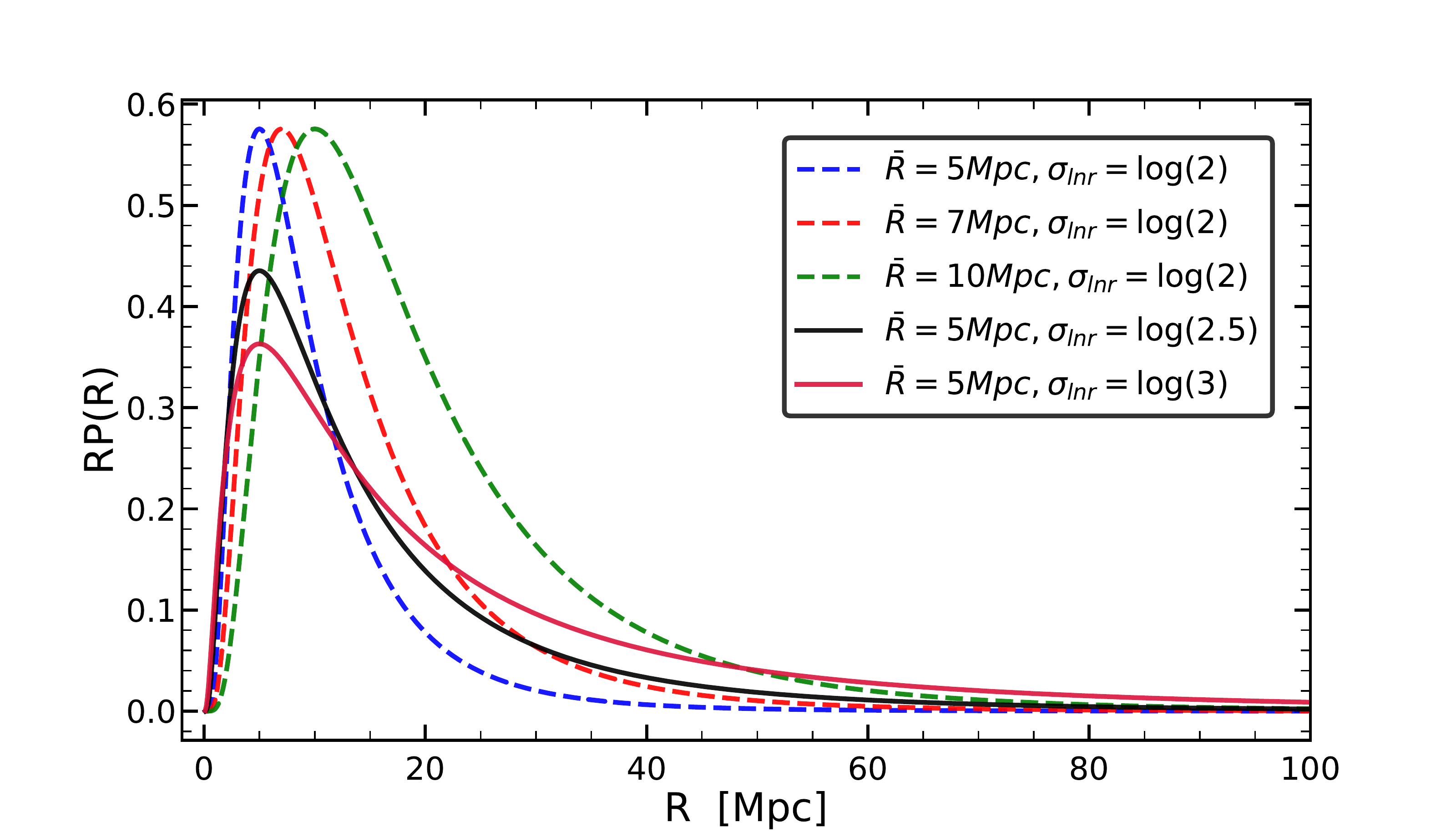}
\caption{ Distribution of bubble radius for different $\bar{R}$ and $\sigma_{lnr}$ as reported.}
\label{rdist}
\end{center}
\end{figure}

\begin{figure}[h]
\begin{center}
\includegraphics[width=1\textwidth]{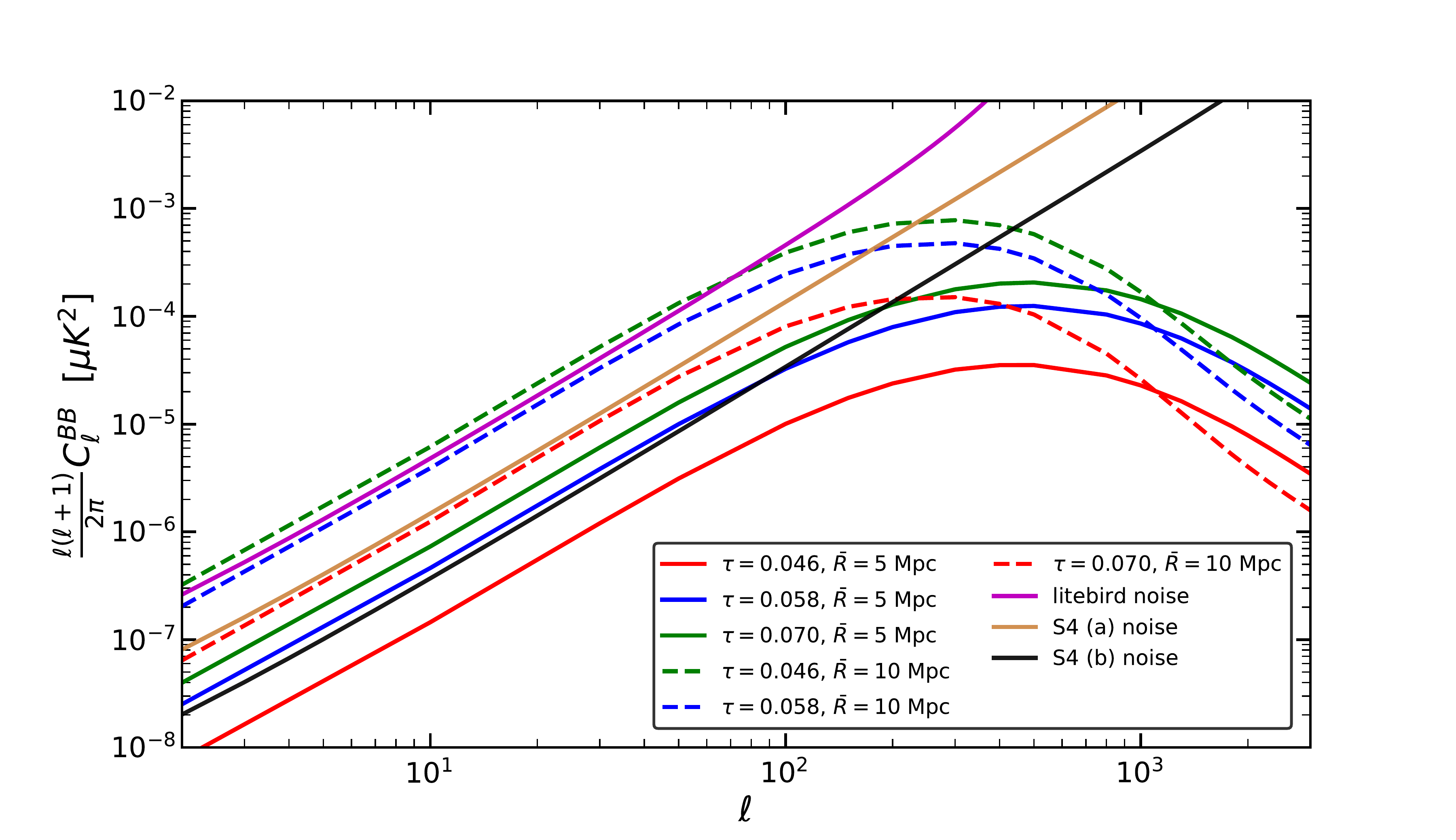}
\caption{Green, Blue and red solid and dashed lines are B mode power spectrum generated due to patchy reionization for different values of $\tau$ and $\bar{R}$ as reported for EERM. Purple, orange and black solid lines are the noise level for the experimental configurations of LiteBIRD, S4 (a) and S4(b) respectively.}
\label{BBpatchy}
\end{center}
\end{figure}

\begin{figure}[h!]
\begin{center}
\includegraphics[width=1\textwidth]{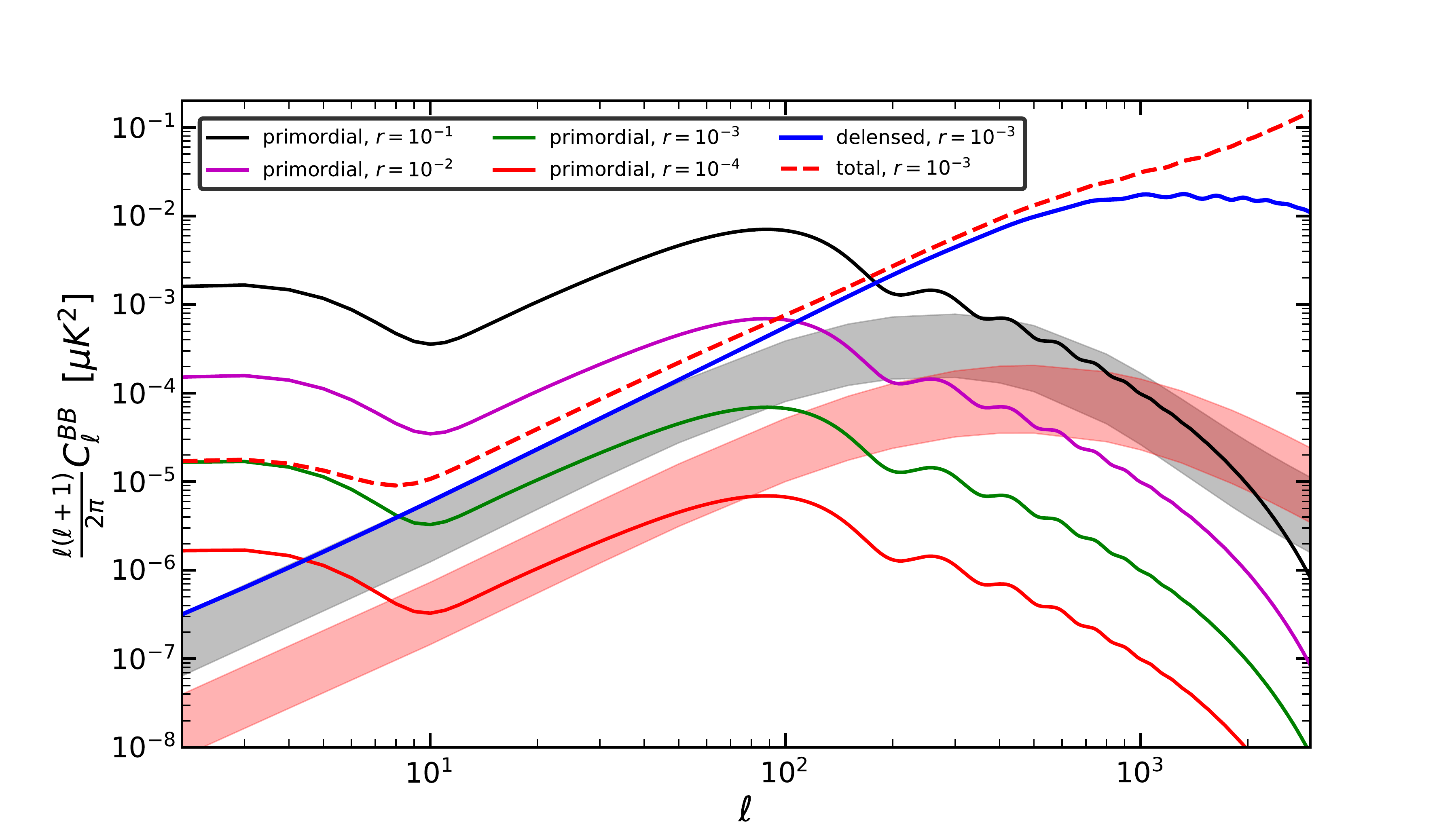}
\caption{Black, magenta, green and red lines are the primordial B mode power spectrum for r=$10^{-1}$, $10^{-2}$, $10^{-3}$,$10^{-4}$ respectively. The blue solid line shows the de-lensed B-mode power spectrum for S4(a) sensitivity and $r=10^{-3}$; the red dashed line adds the primordial signal and instrumental noise to the former.}
\label{BBcontamination}
\end{center}
\end{figure}

\subsection{Patchy Reionization}\label{patchy_reio}

We now focus on morphological aspects of the reionization process, the three most new effects with respect to those discussed previously are the following.

First, since we are considering the inhomogeneous reionization, optical depth $\tau$ will be direction dependent quantity in the CMB sky, and the temperature and polarization anisotropy from the last scattering surface gets screened by a factor of $e^{-\tau(\hat{n})}$ \citep{DvorkinBmode}.

Second, new large scale polarization is generated as CMB local temperature quadrupole is scattered by the ionized regions, thereafter dubbed bubbles.

Third the peculiar motion of the ionized bubbles induces additional temperature fluctuations by KSZ during the epoch of reionization. KSZ effect due to patchy reionization will generate new power on small scales (around $l\approx 3000$) \citep{Battaglia2012, Smith2017} in temperature power spectrum. In this paper we will not consider KSZ contribution as our interest is investigating the polarization anisotropy due to patchy reionization.

We assume the reionization happened by the percolation of ionized HII bubles as well as the the growth in R \citep{Furlanetto2004, Hu2006} inside the neutral intergalactic medium. We consider the size R of ionized bubbles following a log-normal distribution with two free parameters, the characteristic  bubble size $\bar{R}$ (in Mpc) and the standard deviation $\sigma_{lnr}$, given by
\begin{equation}
P(R)=\frac{1}{R}\frac{1}{\sqrt{2\pi\sigma_{lnr}^2}}\exp{\left[-\frac{\{\ln\left(R/\bar{R}\right)\}^2}{2\sigma_{lnr}^2}\right]}.
\end{equation}

A bias b in the dark matter halo distribution will also influence the number density of ionized bubbles, but for simplicity we assume the bubble bias to be not evolving with redshift and independent of the bubble radius (in the redshift range of our interest); in this paper we use $b=6$. In Figure~\ref{rdist} we show the radial distribution of ionized bubbles for EERM where we consider various values of $\bar{R}$ and $\sigma_{lnr}$, be exploited in Section $\ref{forecast}$ to investigate how the reionization model affects the observables of patchy reionization. We stress that the radii distribution of bubbles strongly depends on the overall reionization history. For higher reionization optical depth w.r.t the present case, the distribution of bubbles has been studied numerically considering the complicated evolution of ionized bubbles during the reionization epoch\cite{Su2011}. We will comment more in this point in the conclusion section.

The merging of ionization bubbles creates fluctuations in ionization fraction, $\delta x_e(\bold{\hat{n}},\chi)$ over the mean ionization fraction  $\bar{x}_e(\chi)$ during the inhomogeneous reionization epoch, so that
\begin{equation} \label{deltaxe_eq}
x_e(\hat{\bold{n}},\chi)=\bar{x}_e(\chi)+\delta x_e(\bold{\hat{n}},\chi) ,
\end{equation}
\begin{equation} \label{taunew_eq}
\tau(\bold{\hat{n}},\chi)=c\sigma_T n_{p0}\int_0^\chi \frac{d\chi}{a^2}\left[\bar{x}_e(\chi)+\delta x_e(\bold{\hat{n}},\chi)\right].
\end{equation}
In order to quantify the fluctuations in the ionization fraction, three dimensional power spectrum is expressed by the sum of the 1-bubble and 2-bubble contributions in total power spectrum corresponding to the scales in which r$\ll$R (1b) and r$\gg$R (2b), respectively.
The main assumption for calculating the power spectrum of $\delta x_e$ is that fluctuations in free electron density trace the fluctuations in dark matter density \citep{Furlanetto2004}. We take analytic expressions of 1b and 2b contributions from \citet{Dvorkin2009}, given by:

\begin{equation}\label{p1b_eq}
P^{1b}_{\delta x_e\delta x_e}(k)= x_e(1-x_e)\left[\alpha (k)+\beta(k)\right],
\end{equation}
where the functional forms of $\alpha(k)$ and  $\beta({k})$ are given by
\begin{equation}\label{alphak_eq}
\alpha(k)=\frac{\int dR P(R)[V(R)W(KR)]^2}{\int dR P(R)V(R)},
\end{equation}
\begin{equation}\label{betak_eq}
\beta(k)=\int \frac{d^3\bold{k}^\prime}{(2\pi)^3}P\left(\left | \bold{k}-\bold{k}^\prime \right |\right)\alpha(k^\prime).
\end{equation}
The volume of the bubble is $V(R)=\frac{4}{3}\pi R^3$ and matter power spectrum is P(k). $W(kR)$ is the Fourier transform of the tophat window function with radius R,  given by
\begin{equation}\label{wkr_eq}
W(kR)=\frac{3}{(kR)^3}\left[\sin(kR)-kR\cos(kR)\right].
\end{equation}
We adopt the approximation from \cite{Hu2006} to calculate $\beta(k)$ numerically,  given by
\begin{equation}
\beta({k})=\frac{P(k)\sigma_R^2\int dR P(R )V(R)}{[P^2(k)+\{\sigma_R^2 \int dR P(R)V(R)\}^2]^{1/2}}.
\end{equation}
The 2b contribution is given by
\begin{equation}\label{p2b_eq}
P^{2b}_{\delta x_e\delta x_e}(k)= \left[(1-x_e)\ln(1-x_e)\gamma(k)-x_e\right]^2P(k),
\end{equation}
where $\gamma(k)$ is defined by
\begin{equation}\label{gamma_eq}
\gamma(k)= b \cdot \frac{\int dR P(R)V(R)W(KR)}{\int dR P(R)V(R)}.
\end{equation}
Therefore the total three dimensional power spectrum of $\delta x_e$ can be written as
\begin{equation}\label{ptotal_eq}
P_{\delta x_e\delta x_e}(k)= P^{1b}_{\delta x_e\delta x_e}(k)+P^{2b}_{\delta x_e\delta x_e}(k).
\end{equation}
Now the optical depth power spectrum $C_\ell^{\tau\tau}$ can be constructed from Eq.(\ref{ptotal_eq}). In the flat sky approximation it can be written using Limber approximation as
\begin{equation}\label{cltau_eq}
C_\ell^{\tau\tau}=\sigma_T^2 n^2_{p0}\int  \frac{d\chi}{a^4\chi^2} P_{\delta x_e \delta x_e}\left(\chi, k=\frac{\ell}{\chi}\right).
\end{equation}
Inhomogeneous reionization is also the source of secondary anisotropies in the B mode polarization and the power spectrum can be connected also with $P_{\delta x_e \delta x_e}$ \citep{Hu2000, Mortonson2006} as
\begin{equation}\label{clbbpatchy_eq}
C_\ell^{BB-patchy}=\frac{3\sigma_T^2 n^2_{p0}}{100}\int  \frac{d\chi}{a^4\chi^2}e^{-2\tau({\chi})}Q^2_{rms} P_{\delta x_e \delta x_e}\left(\chi, k=\frac{\ell}{\chi}\right).
\end{equation},

Here $Q_{rms}$ is the r.m.s temperature of the local quadrupole and we used $Q_{rms}=22$ $\mu K$ during the patchy reionization epoch to calculate $C_\ell^{BB-patchy}$. In Figure $\ref{BBpatchy}$ we compare BB spectra from patchy reionization with the foreseen sensitivities from S4 and LiteBIRD. As it can be seen, the signal is at the noise level in both cases, requiring additional steps for increasing the signal to noise ratio, described in the next section. We can see in the figure that S4(a) sensitivity is a bit lower in $\ell \lesssim 200$ than the patchy BB signal both for $\tau=0.070$ and $\tau=0.058$ with $\bar{R}=10$ Mpc. On the other hand, the LiteBIRD sensitivity is a bit higher than patchy BB signal for the same configuration. S4(b) noise level is almost the same with respect to the Patchy BB signal even with $\bar{R}=5$ Mpc. This can be very useful in near future to constrain the morphology of the reionization by observing the patchy B mode signal. We also consider the contamination from patchy reionization to the B-mode power spectrum in figure \ref{BBcontamination}. Adopting the model discussed in the present paper and the $1\sigma$ bounds on $\tau$ from Planck, the contribution from patchy reionization is significant for $r<10^{-3}$.

\section{Reconstruction of $\tau$ along the line of sight}\label{reconstruction}
In Section~\ref{patchy_reio} we discussed how patchy reionization induce additional structure in the observed CMB sky through $\tau$ fluctuations. We consider that $T$ and $(Q\pm iU)$ are the temperature and polarization Stokes parameters before the start of reionization; then, at the end of the reionization, CMB temperature and polarization parameters will be changed as follow \citep{Dvorkin2009}:
\begin{equation}
{T}(\bold{\hat{n}})=T_0(\bold{\hat{n}})+\int \delta \tau T_1(\bold{\hat{n}}),
\end{equation}
\begin{equation}
(Q \pm iU)(\bold{\hat{n}})=(Q\pm iU)_0(\bold{\hat{n}})+ \int \delta \tau (Q\pm iU)_1(\bold{\hat{n}}),
\end{equation}
Fluctuations in optical depth is given by:
\begin{equation}
\delta \tau=\int_{\chi_{start}}^{\chi_{end}}\frac{d\chi}{a^2}\delta x_e(\hat{\bold{n}},\chi).
\end{equation}

Where $\chi_{start}$ and $\chi_{end}$ are the comoving distances to the start and end of reionization respectively. $T_0(\bold{\hat{n}})$ and $(Q \pm iU)_0(\bold{\hat{n}})$ are the contributions coming from recombination and homogeneous reionization and $T_1(\bold{\hat{n}})$ and $(Q \pm iU)_1(\bold{\hat{n}})$ are the contribution from patchy reionization as defined in \citep{Dvorkin2009}.\\

Separately we can write inhomogeneous screening terms as
\begin{equation}
{T}^{scr}=T_0( {\bold{\hat{n}}}) e^{-\delta \tau( {\bold{\hat{n}}})},
\end{equation}
\begin{equation}
{({Q } \pm i {U} )}^{scr}={({Q } \pm i {U} )_0}(\bold{\hat{n}})e^{-\delta \tau(\bold{\hat{n}})},
\end{equation}
accounting for the damping of anisotropies along the line of sight due to patchy reionization.

We now make a parallelism with the formalism applying to CMB lensing. The lensed field $S_{len}(\bold{\hat{n}})$ is related with the unlensed field $S_{unl}(\bold{\hat{n}})$ and the lensing potential $\phi$ as \cite{HuOkamato}
\begin{equation}
S_{len}(\bold{\hat{n}})=S_{unl}(\bold{\hat{n}})+(\nabla\phi) \nabla S_{unl}(\bold{\hat{n}})+O[(\nabla \phi)^2],
\end{equation}
so, it modulates both the CMB polarization and temperature by correlating different Fourier modes. In flat sky approximation, after the lensing reconstruction method \citep{HuOkamato, Dvorkin2009}, this correlation for patchy reionization can be written as
\begin{equation}
\langle S({{\vec{\ell}_1}}) S^\prime ({{\vec{\ell}_2}})\rangle= (2\pi)^2 C_{\ell}^{SS^\prime}\delta({\vec{L}})+ f^{\tau}_{SS^\prime}({\vec{\ell}_1},{\vec{\ell}_2}) [\delta \tau ({\vec{L}})],
\label{tau_eq}
\end{equation}
where $S,S^\prime$ can be any combinations of T,E and B and $\vec{L}=\vec{\ell}_1+\vec{\ell}_2$. We use only EB minimum variance quadratic estimator since it provides the highest signal to noise ratio to patchy reionization \citep{Gluscevic2009, Dvorkin2009}. $f^\tau_{EB}$ for flat sky is given by \citep{Su2011}
\begin{equation}
f^\tau_{EB}=\left(\bar{C}_{\ell_1}^{EE}-\bar{C}_{\ell_2}^{BB}\right)\sin2(\phi_{\ell_1}-\phi_{\ell_2}),
\end{equation}
where $\bar{C}_{\ell_1}^{EE}$ and $\bar{C}_{\ell_2}^{BB}$ are the EE and BB power spectra which include patchy reionization and $\phi_\ell=\cos^{-1}(\hat{\bold{n}}\cdot\hat{\bold{\ell}})$. To satisfy Eq.($\ref{tau_eq}$) we can write minimum variance quadratic estimator of $\tau(\vec{{L}})$ as
\begin{equation}
\widehat{\tau}_{EB}(\vec{{L}})=\widetilde{N}^{\tau}_{EB}(\vec{{L}}) \int \frac{d^2{{\vec{\ell}_1}}}{(2\pi)^2}\left[E({{\vec{\ell}_1}})B({{\vec{\ell}_2}})   \right] F^\tau_{EB}({{\vec{\ell}_1}},{{\vec{\ell}_2}}),
\end{equation}
where $\widetilde{N}^{\tau}_{EB}({{\vec{L}}})$ is the zeroth order bias for $\tau$ reconstruction which can be thought in analogy to the $N_{EB}^{(0)}$ bias for lensing  potential reconstruction, given by
\begin{equation}
\widetilde{N}^{\tau}_{EB}({{\vec{L}}})= \left[ \int \frac{d^2{{\vec{\ell}_1}}}{(2\pi)^2}  f^{\tau}_{EB}({\vec{\ell}_1},{\vec{\ell}_2}) F^\tau_{EB}({{\vec{\ell}_1}},{{\vec{\ell}_2}})\right]^{-1}.
\end{equation}
The CMB instrumental noise power spectrum for a Gaussian symmetric beam can be written as
\begin{equation}\label{noise_eq}
N_\ell^{P}=\Delta_P ^2\exp\left[\frac{\ell(\ell+1)\Theta^2_f}{8\ln2}\right],
\end{equation}
where $\Delta_P$ is the noise of the detector for polarization in $\mu K$-arcmin; This is $\sqrt{2}$ times bigger than the detector noise for temperature and $\Theta_{f}$ is the full width half maxima of the beam in arcmin units.

In order to minimize the variance of $\langle \widehat{\tau}_{EB}({{\vec{\ell}_1}})\widehat{\tau}_{EB}({{\vec{\ell}_2}}) \rangle$, the optimal form of filter $F(\vec{{\ell_1}},\vec{{\ell_2}})$ for EB estimator is given by
\begin{equation}
F^\tau_{EB}(\vec{{\ell_1}},\vec{{\ell_2}})=\frac{ f^{\tau}_{EB}({\vec{\ell_1}},{\vec{\ell_2}})}{(C_{\ell_1}^{EE}+N_{\ell_1}^{EE})(C_{\ell_2}^{BB}+N_{\ell_2}^{BB})} ,
\end{equation}

So that the expectation value of the estimator for $\tau$ will reduce to
\begin{equation}
\langle \widehat{\tau}_{EB}({{\vec{\ell}_1}})\widehat{\tau}_{EB}({{\vec{\ell}_2}}) \rangle=(2\pi)^2\delta({{\vec{\ell}_1}},{{\vec{\ell}_2}})\left[ C_L^{\tau\tau}+\widetilde{N}^{\tau}_{EB}({{\vec{L}}})\right].
\end{equation}
We will use this technique in the next section to reconstruct the power spectrum of $\tau$ using EERM prescription.

\section{Detectability of the reionization angular power spectrum through $\tau$ reconstruction in S4 CMB experiments}\label{forecast}

In this section we discuss the capabilities of future CMB polarization experiments to detect the angular pattern of $\tau$. In order to do so we have modified the publicly available {\tt Lenspix}\footnote{cosmologist.info/lenspix} algorithm \citep{Lewis2011} to reconstruct $\tau$ instead of the GL potential $\phi$ following the algebra discussed in the previous Section. We focus on Eq.(\ref{cltau_eq}) in order to calculate the $C_\ell^{\tau\tau}$ for our reionization model, EERM.

\begin{figure*}[h!]
\begin{center}
\includegraphics[width=1\textwidth]{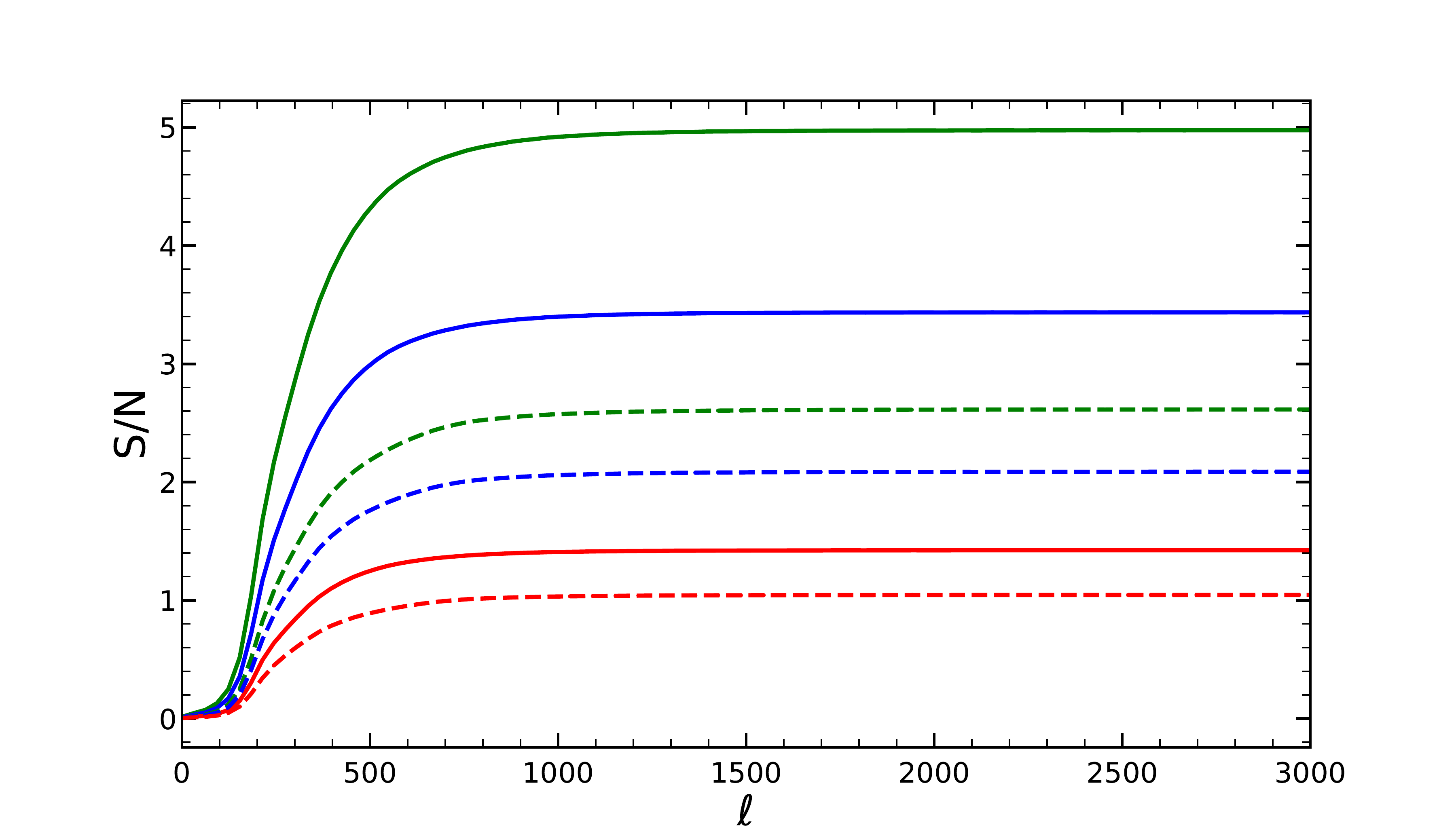}
\caption{Solid lines represents the cumulative SNR for EERM and dashed lines for the tanh model with $\tau$= 0.070 (green), 0.058 (blue) and 0.046 (red) respectively.}
\label{fig:cltau_astro}
\end{center}
\end{figure*}

The uncertainty in the power spectrum of $\tau$ comes from both the astrophysical modelling of the source of reionization and the uncertainty due to the modelling of reionization morphology. The first source of uncertainty is connected with the mean (sky averaged) optical depth $\bar{\tau}$, escape fraction $f_{esc}$ and the source of reionization, which in our case is the star formation rate functions of high redshift galaxies. The morphology of reionization depends on the the bias factor, the mean radius of bubbles $\bar{R}$ and also the spread of distribution of bubbles radii which is quantified by $\sigma_{lnr}$ for a log normal distribution. We will consider the effects of $\bar{R}$, $\bar{\tau}$, and $\sigma_{lnR}$ uncertainties on the power spectrum of $\tau$  for EERM, implementing a minimum variance quadratic estimator.

\begin{figure*}[t]
\begin{center}
\includegraphics[width=1\textwidth]{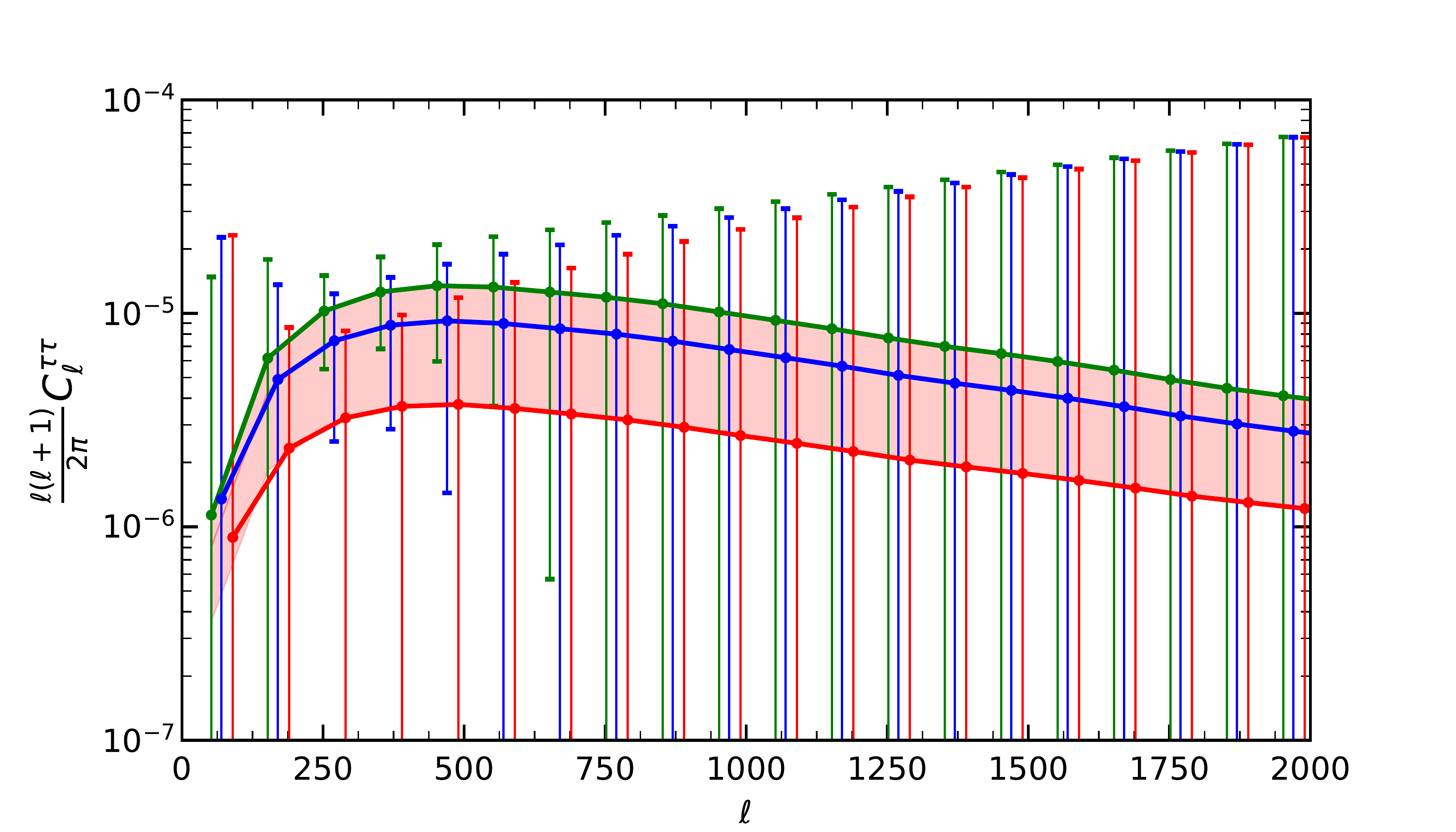}
\caption{Green, blue and red solid lines represent the $\tau$ angular power spectra corresponding to $\bar{\tau}= 0.07, 0.058$ and $0.044$, respectively with $\bar{R}=5$ Mpc and $\sigma_{lnr}=\ln(2)$. We bin the each spectrum with $\Delta \ell=100$ and show the error bars accordingly. The error bars are located at $l_{error}=(\ell_{min}+\frac{(2n-1)\Delta \ell}{2})$, where $\ell_{min}$ is the value of minimum $\ell$ and $n$ is the bin index; we fix $\ell_{min}$ =2 (green curve), 20 (blue curve), 40 (red curve). Shaded pink area represents due to the $1\sigma$ error of the $\bar{\tau}$ measurement by Planck \citep{Planckxlvii}.}
\label{fig:cltau_astro}
\end{center}
\end{figure*}

\begin{figure}[t]
\begin{center}
\includegraphics[width=1\textwidth]{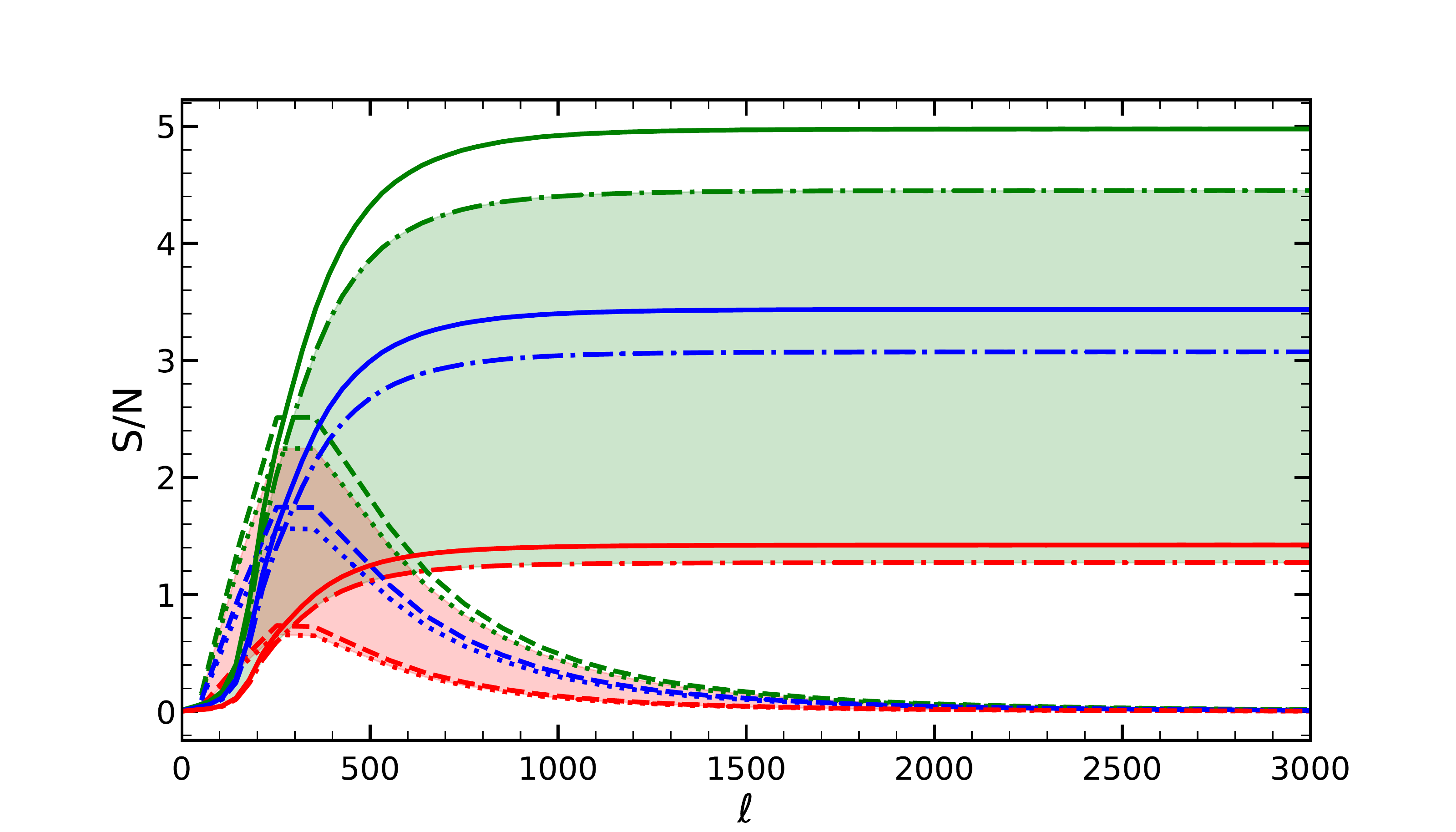}
\caption{The binned and integrated SNR for the cases in Figure $\ref{fig:sn_astro}$. Solid green, blue and red curve represent the integrated SNR for S4 sensitivity with $f_{sky}=0.5$ for $\bar{\tau}=0.07,0.058,0.044$, respectively, while the dashed dotted curves are for $f_{sky}=0.4$. Dashed green, blue and red lines represent the SNR in each bin for $f_{sky}=0.5$, while dotted lines are for $f_{sky}=0.4$.}
\label{fig:sn_astro}
\end{center}
\end{figure}

\begin{figure*}[h!]
\begin{center}
\includegraphics[width=1\textwidth]{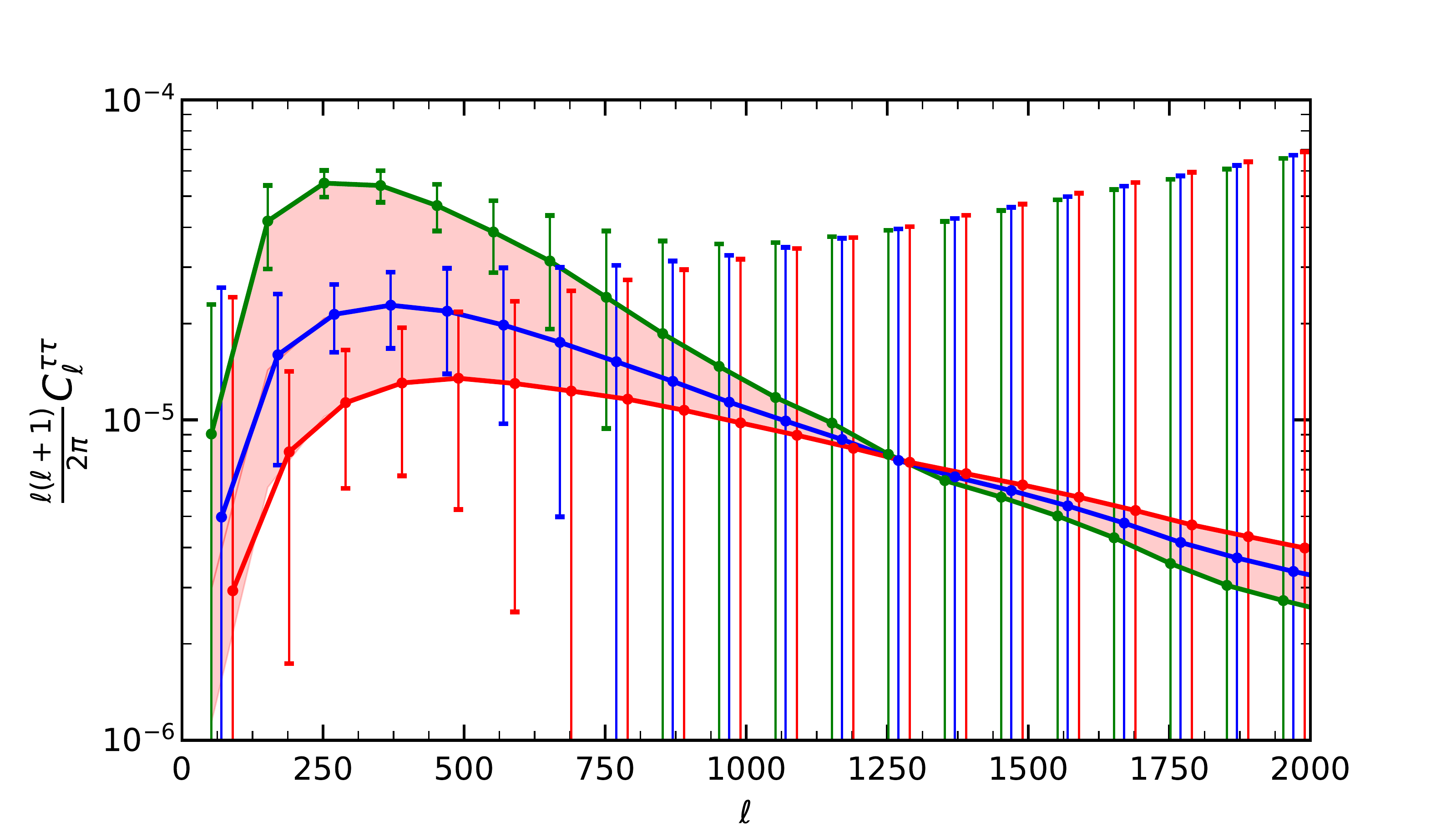}
\caption{Green, blue and red solid lines are the $\tau$ spectra with $\bar{R}=10,7,5$ Mpc, respectively, with $\sigma_{lnr}=\ln2$ for $\bar{\tau}=0.070$. The binning scheme is as in the previous figures.}
\label{fig:cltau_rbar}
\end{center}
\end{figure*}

\begin{figure}[h!]
\begin{center}
\includegraphics[width=1\textwidth]{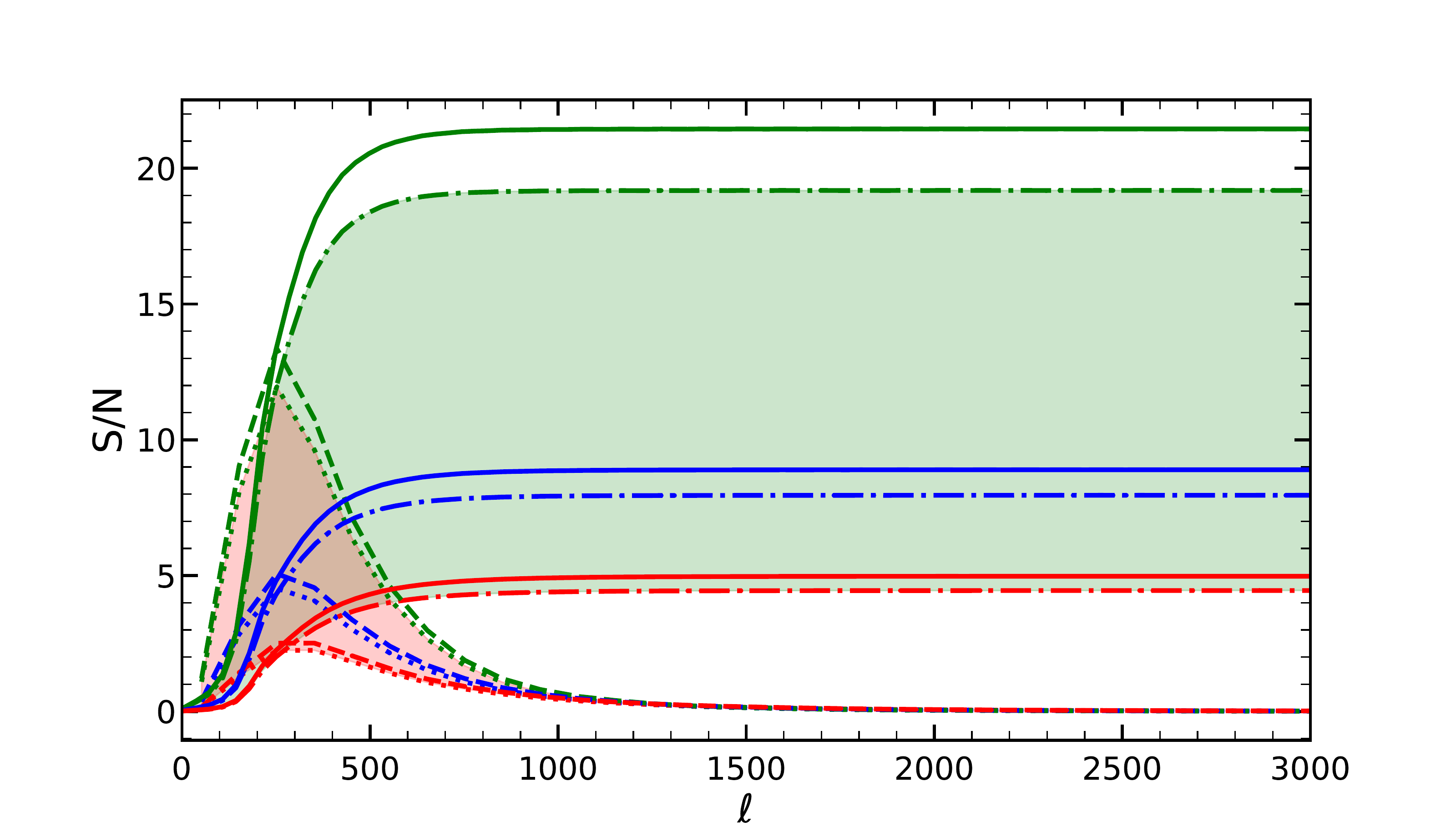}
\caption{Binned and integrated SNR for the cases in Figure $\ref{fig:cltau_rbar}$. Solid green, blue and red curve represent the integrated SNR for S4 sensitivity with $f_{sky}=0.5$ for three optical depths ,$\bar{\tau}=0.07,0.058,0.044$, respectively; dashed dot curves are for $f_{sky}=0.4$ with the same configuration. Dashed
green, blue and red lines are SNR in each bin for $f_{sky}=0.5$, while dotted lines for $f_{sky}=0.4$.}
\label{sn_rbar}
\end{center}
\end{figure}
														
\begin{figure*}[h!]
\begin{center}
\includegraphics[width=1\textwidth]{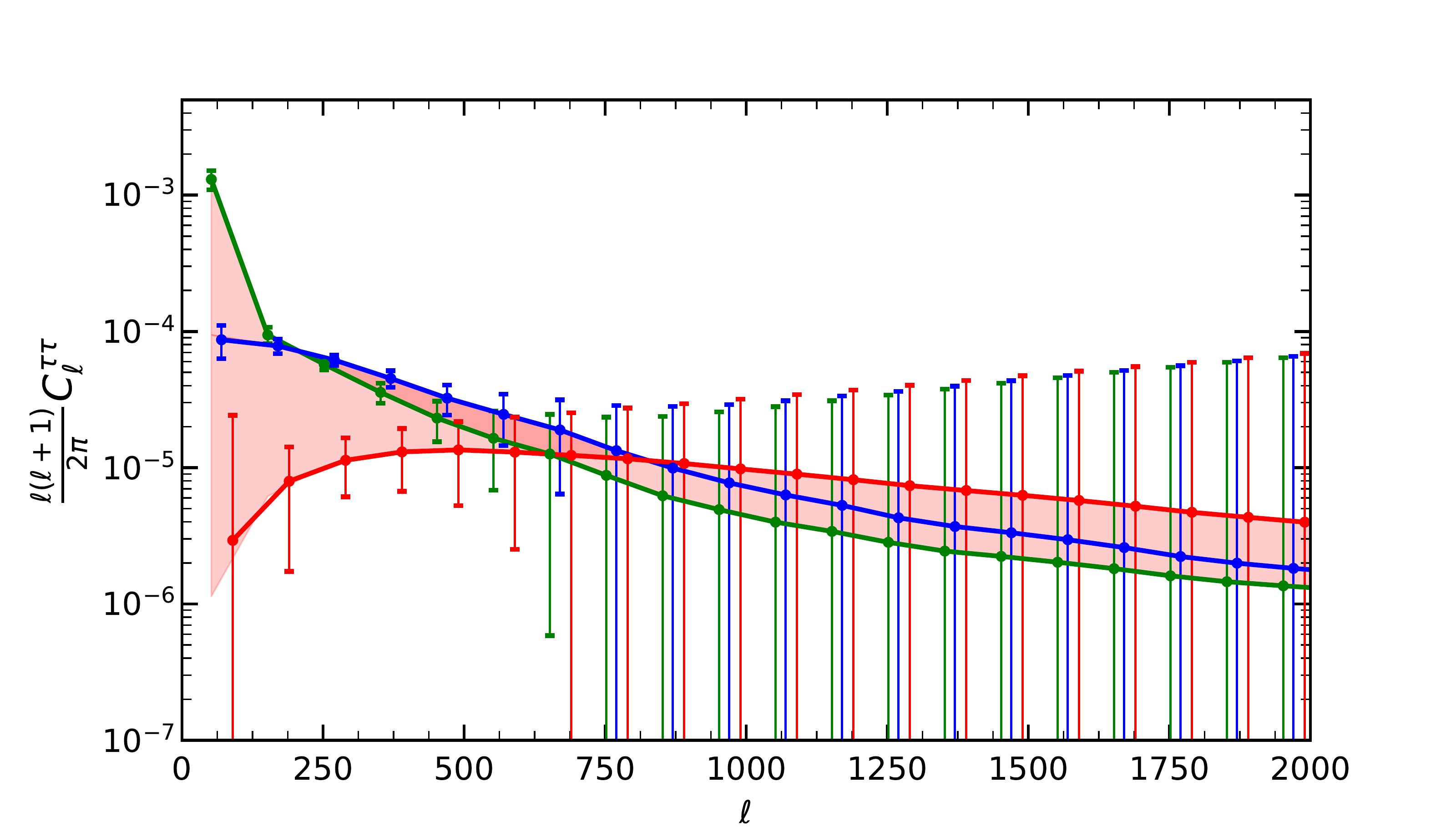}
\caption{Green, blue and red solid lines are the $\tau$ spectra for $\sigma_{lnr}=ln3,ln2.5, ln3 $, respectively, with $\bar{R}=5$ Mpc for $\bar{\tau}=0.070$. The binning scheme is as in the previous figures.}
\label{fig:cltau_sigma}
\end{center}
\end{figure*}		
			
\begin{figure}[h!]
\begin{center}
\includegraphics[width=1\textwidth]{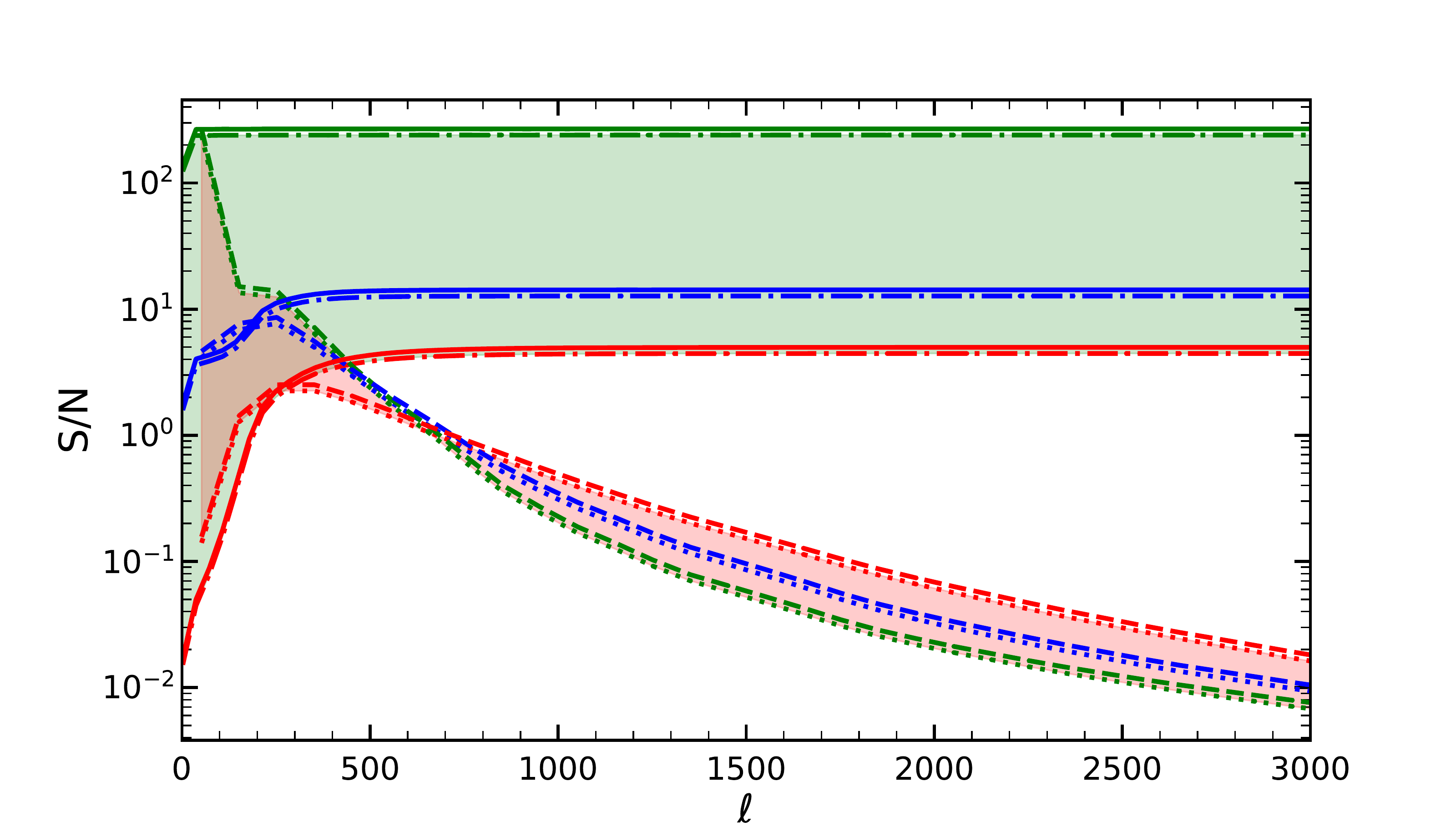}
\caption{Binned and integrated SNR for the case in Figure $\ref{fig:cltau_rbar}$. Solid green, blue and red curves represent the integrated SNR for S4 sensitivity with $f_{sky}=0.5$ for three optical depths $\bar{\tau}=0.07,0.058,0.044$, respectively, and dashed dotted curves are for $f_{sky}=0.4$ with the same configuration. Dashed green, blue and red lines are SNR in each bin for $f_{sky}=0.5$, while dotted lines for $f_{sky}=0.4$.}
\label{sn_sigma}
\end{center}
\end{figure}

We calculate the Signal to Noise Ratio (SNR) to quantify the detectability of the patchy reionization:
\begin{equation}
\left(\frac{S}{N}\right)=\left[\frac{f_{sky}}{2}\sum_{l_{min}}^{l_{max}}(2\ell+1)\left(\frac{C_\ell^{\tau\tau}}{\widetilde{N}_\ell^{\tau\tau}}\right)^2\right]^{1/2}.
\end{equation}
Here $f_{sky}$ is the observed sky fraction for a particular experiment. We mainly focus on instrumental specifications corresponding to S4, for which we use $\Theta_{f}=1$ arcmin and $\Delta_T=1$ $\mu K$-arcmin, i.e $\Delta_P=\sqrt{2}$ $\mu K$-arcmin. We found that SNR for $\tau$-reconstruction does not strongly depend on $\Theta_f$ as SNR changes a factor $\approx$ 3 when $\Theta_f$ increased from $1^\prime$ to $5^\prime$ but SNR increases by a factor of $\approx$ 80 by the decrease of $\Delta_T$ from 2 $\mu K$-arcmin to 0.5 $\mu K$-arcmin. The EERM, featuring a more extended patchy reionization history, tend to yield higher SNR with respect to the tanh model.

\subsection{Dependence on $\bar{\tau}$}

In Figure $\ref{fig:cltau_astro}$ we compute $C_l^{\tau\tau}$ for three different optical depths, corresponding to the central value and $\pm 1\sigma$ of the most recent measurements from Planck, and with fixed reionization parameters $\bar{R}=5$ $Mpc$ and $\sigma_{lnr}=\ln(2)$. For $\bar{\tau}=0.044$, the errors are too large at all scales for detectability, even for $f_{sky}=0.4$. For $\bar{\tau}=0.070$ and $0.058$, an S4 experiment would be capable of probing $\tau$ fluctuations in multipole range $170\lesssim\ell\lesssim 720$ and $190\lesssim\ell\lesssim 630$,  respectively, roughly corresponding to an interval from 1 to 1/3 of a degree. For a fixed detector sensitivity and beam resolving power, a measurement of the reionization morphology through $C_\ell^{\tau\tau}$ is simply proportional to the value of mean optical depth $\bar{\tau}$. If LITEBIRD \citep{Matsumura2014} achieves its sensitivity goal of $\sigma(\bar{\tau})=0.002$, then this would provide an important prior for detecting and interpreting the inhomogeneous reionization signal.

In Figure $\ref{fig:sn_astro}$, we plot the binned and integrated SNRs. Depending on the values of $\bar{\tau}$, the CMB S4 specification should allow a $1.4- 5 \sigma$ detection of the signal. For $\tau=0.44$, the SNR$<$1 at all $\ell$s in the binned spectra. For $\bar{\tau}=0.7$ and $\bar{\tau}=0.58$, S/N is greater than 1 in $152\lesssim\ell\lesssim 752$ and $170\lesssim\ell\lesssim 570$, respectively.

\subsection{Dependence on $\bar{R}$ and ${\sigma_{lnr}}$}

The size and merger history of ionized bubbles is poorly understood. We consider the variation of reionization bubble radius $\bar{R}$ and standard deviation of the log normal distribution of $\sigma_{lnr}$.

If the bubbles are larger ($\bar{R}\approx 10$ Mpc), then the CMB S4 can probe $\tau$ spectra in wide range of $\ell$s as error bars are small in $52\lesssim\ell\lesssim 852$ compared with the signal level. Whereas for $\bar{R}=7$ Mpc and $\bar{R}=5$ Mpc error bars are considerably larger than the signal level in the range of multipoles $70\lesssim\ell\lesssim 770$ and $90\lesssim\ell\lesssim 690$, respectively.

In Figure $\ref{sn_rbar}$ we can see that as $\bar{R}$ increases, the integrated SNR also increases. In the range of $5\lesssim \bar{R} \lesssim 10$ Mpc, the signature of patchy reionization can be measured with $4.3\sigma$ to $19.2\sigma$ by covering $40\%$ of the sky and $4.8\sigma$ to $21.3\sigma$ by observing $50\%$ of the sky.

In Figure $\ref{fig:cltau_sigma}$ we can see that for $\sigma_{lnr}=\ln3$ and $\sigma_{lnr}=\ln2.5$, $C_\ell^{\tau\tau}$ can be measured in the range $52\lesssim\ell\lesssim 652$ and $70\lesssim\ell\lesssim 670$.
In the Figure $\ref{sn_sigma}$, the integrated SNR reaches 267 which is much larger than the variation of $\bar{R}$ and $\bar{\tau}$. Thus, due to the variation of $\sigma_{lnr}$ in between $\ln2$ to $\ln3$,the S4 specifications would allow to detect patchy reionization with a confidence ranging from $4.8\sigma$ to $267\sigma$.

\clearpage
\section{Summary and Outlook}\label{sec|summary}
Cosmic reionization should be inhomogeneous with large ionized regions embedded in a shrinking neutral phase. This inhomogeneous reionization should produce a detectable signal.  With its unprecedented sensitivity, CMB S4 should be capable of detecting the signal at a minimum, constraining  the physics of reionization.

In this paper, we construct an ionizing background history that is based on recent astrophysical observations. We have adopted a simple parametrization of the size distribution of the early ionized regions; specifically we use a log-normal distribution in radius, characterized by a mean radius and a spread. The sky pattern of the reionization in such a picture is commonly known as patchy.

Following earlier works, we have implemented a procedure for the extraction of the patchy reionization signal analogue to the one exploited for CMB lensing and evaluated the signal to noise ratio achievable by CMB experiments reaching the S4 capabilities. The angular power spectrum is distributed broadly around the degree scale, with a long tail at higher multipoles. We have studied the amplitude of the signal as a function of the overall properties of the bubbles, and in particular on the abundance of the largest ones. If the bubble distribution has a tail with bubble size extending to tens of Mpc, then CMB S4 SNR could be in the many tens to hundreds.

Because our predicted signal is one order of magnitude larger than that estimated by \citet{Su2011}, lensing noise contamination should be a less relevant issue for us. Since the lensing behaviour is known for a given set of cosmological parameters, we can remove its contamination from our signal. These results demonstrate the capability of future CMB experiments to detect patchy reionization contribution to the CMB anisotropies.

\acknowledgments{We thank the referee for helpful comments and suggestions. AR would like to thank Luigi Danese, Davide Poletti and Arnab Chakraborty for useful discussion throughout this project. This work partially supported by PRIN MIUR 2015 `Cosmology and Fundamental Physics: illuminating the Dark Universe with Euclid', PRIN INAF 2014 `Probing the AGN/galaxy co-evolution through ultra-deep and ultra-high-resolution radio surveys', the MIUR grant `Finanziamento annuale individuale attivita base di ricerca' and by the RADIOFOREGROUNDS grant (COMPET-05-2015, agreement number 687312) of the European Union Horizon 2020 research and innovation program.}

\bibliographystyle{apj}
\bibliography{mybib}

\end{document}